\documentclass[twocolumns]{emulateapj}

\usepackage{color, colortbl}
\usepackage{url}
\usepackage{natbib}
\usepackage{amssymb}
\usepackage{amsmath}
\def\be{\begin{eqnarray}}
\def\ee{\end{eqnarray}}
\usepackage{comment}
\usepackage{color}
\usepackage{graphicx,subfigure}
\usepackage{enumerate}
\usepackage{rotating}
\usepackage{longtable}
\renewcommand{\arraystretch}{1.5}

\definecolor{pink}{RGB}{255, 20, 147}

\definecolor{carminered}{rgb}{1.0, 0.0, 0.22}

\definecolor{amber}{rgb}{0.95, 0.8, 0.2}

\definecolor{blue}{rgb}{0.01, 0.01, 0.98}

\definecolor{byzantine}{rgb}{0.74, 0.2, 0.64}

\definecolor{amethyst}{rgb}{0.6, 0.4, 0.8}

\definecolor{blue-violet}{rgb}{0.54, 0.17, 0.89}

\definecolor{blue-violet}{rgb}{0.54, 0.17, 0.89}

\definecolor{blue-violet}{rgb}{0.54, 0.17, 0.89}

\definecolor{comment}{RGB}{166, 38, 164}

\shorttitle{Search for radio remnants of nearby off-axis GRBs}

\shortauthors{Grandorf, McCarty, et al.}

\begin{document}

\title{Search for radio remnants of nearby off-axis Gamma-Ray Bursts in a sample of \textit{Swift}/BAT events}
\author{C.~Grandorf\altaffilmark{1,$\star$}, J.~McCarty\altaffilmark{1,$\dagger$}, P.~Rajkumar\altaffilmark{1}, H.~Harbin\altaffilmark{1},
K.H.~Lee\altaffilmark{2,\S},
% IB: I moved Lee two up so students are first and then senior authors, please feel free to reshuffle
A.~Corsi\altaffilmark{1},
I.~Bartos\altaffilmark{2},
Z.~M\'arka\altaffilmark{3},
A.~Balasubramanian\altaffilmark{1},
S.~M\'arka\altaffilmark{4}}
\altaffiltext{1}{Department of Physics and Astronomy, Texas Tech University, Box 1051, Lubbock, TX 79409-1051, USA}
\altaffiltext{2}{Department of Physics, University of Florida, Gainesville, FL 32611-8440, USA}
\altaffiltext{3}{Columbia Astrophysics Laboratory, Columbia University in the City of New York, New York, NY 10027, USA}
\altaffiltext{4}{Department of Physics, Columbia University in the City of New York, New York, NY 10027, USA}

\email{$^{*}$Connor.Grandorf@ttu.edu, ConnorGrandorf@my.unt.edu \\ $^{\dagger}$jmccarty@mit.edu \\ $^{\S}$kyunghwanlee@ufl.edu}

\begin{abstract}
\label{abstract}
The multi-messenger discovery of gravitational waves (GWs) and light from the binary neutron star (NS) merger GW170817, associated with Gamma-Ray Burst (GRB) 170817A and kilonova AT2017gfo,  has marked the start of a new era in astrophysics.  
GW170817 has confirmed that binary NS mergers are progenitors of at least some short GRBs. The peculiar properties of the GRB\,170817A radio afterglow, characterized by a delayed onset related to the off-axis geometry, have also demonstrated how some nearby short GRBs may not be identified as such with standard short-timescale electromagnetic follow-up observations.  Building upon this new information, we performed late-time radio observations of a sample of four short GRBs with unknown redshift and no previously detected afterglow in the \textit{Swift}/BAT sample in order to identify nearby ($d_L\lesssim 200$\,Mpc) off-axis GRB candidates via their potential late-time radio signatures. We find a previously uncatalogued radio source within the error region of GRB\,130626 with a $3-6$\,GHz flux density consistent with a NS radio flare at a distance of $\sim100$\,Mpc. However, an origin related to a persistent radio source unrelated to the GRB cannot be excluded given the high chance of false positives in error regions as large as those considered here.   Further radio follow-up observations are needed to better understand the origin of this source.
\end{abstract}


\section{Introduction}
\label{intro}
GW170817, the first binary neutron star (NS) merger detected by the LIGO and Virgo gravitational wave (GW) detectors, was accompanied by an electromagnetic (EM) counterpart observed in all bands of the EM spectrum \citep[][and references therein]{Abbott2017b,Abbott2017a,Abbott2017c}. Located at $\sim40$\,Mpc \citep[e.g.,][]{Blanchard2017,Coulter2017}, it was associated with the closest and most sub-energetic short $\gamma$-ray burst (GRB) we know of, GRB\,170817A \citep{Abbott2017b,Fong2017,Goldstein2017,Savchenko2017}. GW170817 also provided the first direct evidence of a kilonova, a quasi-thermal UV-optical-IR transient  powered by the radioactive decay of heavy elements \citep[dubbed AT2019gfo; e.g.][]{Arcavi2017,Chornock2017,Cowperthwaite2017,Coulter2017,Drout2017,Kasliwal2017,Nicholl2017,Soares-Santos2017,Valenti2017}. The delayed afterglow of GRB\,170817A, discovered in X-rays and radio $\approx 9$ and $\approx 15$ days after the merger, respectively, was unusual \citep{Abbott2017c,Hallinan2017,Haggard2017,Troja2017}. Indeed, we now know that GW170817 also represents the very first secure observation of an off-axis GRB  \citep{Alexander2017,Margutti2017,Corsi2018,Dobie2018,Margutti2018,Mooley2018a,Mooley2018b,Mooley2018c,Troja2018a,Ghirlanda2019,Hajela2019,Troja2019,Makhathini2020}. The overall picture that has emerged via extended radio follow-up is that of a structured relativistic outflow with an energetic jet core of half opening angle $\theta_{\rm j}\sim5^\circ$, accompanied by slower, less-energetic wings on larger angular scales (tens of degrees) from the jet axis \citep{Granot2018,Lazzati2018,Mooley2018c}. This structured jet is substantially different from the top-hat jet of uniform brightness typically assumed to model EM observations of cosmological GRBs \citep[e.g.,][]{Berger2014,vanEerten2012}. 

The new evidence for structured jets brought by GW170817/GRB\,170817A, together with the surprising proximity of this event, have raised the question of whether the known sample of short GRBs could be hiding a nearby population of GW170817-like events. In fact, GW170817-like afterglows could have been missed by the typical localization strategy based on rapid X-ray and radio follow-up observations of GRBs. While a kilonova component similar to AT2017gfo would be well above the sensitivity of most ground-based telescopes up to distances of $\approx 200$\,Mpc, a kilonova fainter or redder than AT2017gfo \citep[e.g.,][]{Metzger2014} could have been easily missed in past searches as well. On the other hand, the late-time radio flares predicted to arise by the interaction of the fastest kilonova ejecta with the circum-binary medium could still be detectable years after the GRB \citep[e.g.,][]{Nakar2011,Hotokezaka2013,Hotokezaka2016,Hotokezaka2018,Nakar2019,Liu2020,Margalit2020}. 

The above considerations have motivated a variety of studies aimed at hunting for nearby GRBs in the currently known sample of bursts, and at constraining their rate using potential late-time radio signatures   \citep{Gupte2018,Fong2016,Mandhai2018,Troja2018b,Yue2018,Bartos2019,Klose2019,vonKienlin2019,Matsumoto2020,Ricci2020,Schroeder2020}. Other studies have also looked at the possibility of discovering late-time radio flares by kilonova ejecta using radio survey data alone \citep{Law2018,Lee2020}. 
Among the above studies, \citet{Mandhai2018} did not find any robust evidence for a population of local short GRBs and constrained their all-sky rate to $\lesssim 4$\,yr$^{-1}$ within 200\,Mpc. \citet{Gupte2018} instead concluded that up to 10\% of the total short GRB sample could be nearby. More recently, \citet{Dichiara2020} have cross-correlated short GRB positions with a catalogue of nearby galaxies, and found four possible associations at $100$\,Mpc$\,\lesssim d_L\lesssim 200$\,Mpc. \citet{Ricci2020} followed-up these GRBs in the radio and set constraints on their potential radio flares based on radio non-detections.

Here, we focus on the sample of short GRBs presented in \citet{Bartos2019} as worth monitoring in the radio to uncover potential nearby events. The radio remnants of these nearby GRBs were estimated to still be potentially detectable by the Karl G. Jansky Very Large Array (VLA) if located within 200\,Mpc \citep[the advanced LIGO horizon distance for binary NS mergers when reaching design sensitivity; ][]{Abbott2018}, and if their circum-merger medium is sufficiently dense.
In Section \ref{sample} we briefly summarize the criteria we followed to select our GRB sample. We refer the reader to \citet{Bartos2019} for a more detailed description of the sample selection.  In Section \ref{VLAobs} we describe our radio follow-up observations and data reduction. In Section \ref{results} we discuss our results, and in Section \ref{sec:conclusion} we summarize and conclude.

\section{GRB sample selection}
\label{sample}
To identify nearby GRB candidates in the \textit{Swift} sample, we used the following criteria: 
\begin{enumerate}
\item {\bf Short GRBs --- } We only considered GRBs that were identified in the literature as short based on their duration ($T_{90}$) and spectral hardness \citep{Kouveliotou1993}.
\item {\bf Lack of an afterglow detection --- } We considered only GRBs with no early afterglow detections so as to specifically target potential off-axis events whose afterglow would become visible beyond the typical time frame of X-ray and optical follow-up observations. 
\item {\bf Accurate gamma-ray localization --- } Without a detected afterglow, a GRB is typically localized solely from its $\gamma$-ray emission and thus localization errors are larger than the arcsec localization radii of GRBs with early afterglow detections. In order to minimize the time required for deep radio follow-up, we thus restricted our selection to GRBs that had been observed with the Burst Alert Telescope (BAT) on-board the Neil Gehrels \textit{Swift} Observatory \citep{Gehrels2004}. Bursts in this sample typically have 90\% localization error radii and systematic uncertainties smaller than half the full-width at half maximum of the VLA primary beam at $6$\,GHz \citep{Lien2016}.  
\item {\bf Kilonova constraints --- } We additionally excluded GRBs with sufficiently sensitive and timely optical follow-up observations that would have detected the kilonova emission from the binary merger that produced the GRB if (i) the merger was within 200\,Mpc and (ii) the kilonova emission had the same luminosity and temporal profile as was observed for GW170817.
\item {\bf VLA observability and detectability ---} To ensure observability of GRBs in our sample with the VLA we excluded GRBs with declination below $-40^\circ$ as these are too South. Additionally, we excluded GRBs in the declination range $[-5^\circ,15^\circ]$, to avoid Radio Frequency Interference (RFI) associated with satellites in the Clarke Belt. Finally, we only considered events for which a realistic circum-merger density of $\lesssim1$\,cm$^{-3}$, and two observations in 2019 and 2020 with a reasonable VLA observing time, could potentially result in detections \citep[see][for more details]{Bartos2019}.
\end{enumerate}

None of the GRBs in our sample here have cataloged galaxies within 200\,Mpc that overlap with their localization (the catalog completeness on this distance scale is only 40\% at present \citealt{Bartos2019}).  We note that the GRBs included here are the subset of those presented in \citet{Bartos2019} for which two late-time VLA observations were carried out via our programs (VLA/19A-184 and VLA/20A-239; PI: Bartos). 

\begin{table*}
\begin{center}\begin{footnotesize}
\caption{Sensitivity reached in our C-band (nominal central frequency of 6\,GHz) and S-band (nominal central frequency of 3\,GHz) observations of the GRBs in our sample. We also indicate with $\Delta T$ the epoch (days since GRB trigger) of our observations. Only GRBs for which a candidate radio source was found in their corresponding BAT error circle were re-observed in C-band (see Table \ref{tb:Table}).\label{tb:TableRMS}}
\begin{tabular}{ccclc}
\hline
\hline
GRB Field Name & Freq. & rms  & UT Date & $\Delta T$ \\
&  (GHz) & ($\mu$Jy) & & (days)\\
\hline\hline
130626 & 6.2 & 7.3 & 08 Jul 2019 & 2203\\
'' & 5.2 & 15 &  29 Mar 2020 & 2468\\
'' & 2.9 & 12 & 24 Jul 2019 & 2219\\
141205 & 6.2 &6.5 &02 Apr 2019 &1580\\
'' & 2.8 & 10 & 14 May 2019 & 1621\\
151228 & 6.2 &8.6 &20 Jun 2019 & 1270\\
'' & 6.2 & 10 & 21 May 2020 & 1606\\
'' & 2.9 & 15 & 17 Jul 2019 & 1297\\
170112 & 6.2 &7.9 &30 Mar 2019 & 808\\
'' & 6.3 & 20 & 27 Feb 2020 & 1142\\
'' & 2.8 & 20 & 16 May 2019 & 855\\
\hline
\end{tabular}
\end{footnotesize}\end{center}\end{table*}

\begin{table*}
\begin{center}\begin{footnotesize}
\caption{We report information about any radio source found within the BAT error region of GRBs in our sample. Columns are, from left to right: GRB field name, R.A. and Dec. of the identified radio source(s) from our VLA images; object class of the closest NED counterpart within $1\arcmin$ radius of the VLA position; Modified Julian Date (MJD) of our VLA observations; epoch in days since the GRB trigger time; VLA nominal central frequency; VLA peak flux density; offset between the source location as measured in our VLA images and that reported in NED; VLA position error. \label{tb:Table}}
\begin{tabular}{ccccccccccc}
\hline
\hline
GRB Field Name &  R.A.\,Dec. (VLA) &  Class& Epoch & $\Delta T$ & $\nu$ & $F_{\nu}$ (VLA) & Offset & Pos.Err. (VLA) \\
& (hh:mm:ss\,deg:mm:ss) & & (MJD) & (day) & (GHz)& (mJy) & (\arcsec) & (\arcsec)\\
\hline
\hline
 $130626$ & 18:12:34.00 $-$09:30:02.6 & IrS & 58672.14 & 2203 & 5.2%6.2 
 & $0.103\pm0.016$%$0.097 \pm 0.010$ 
 & 4.1
 & 0.10 \\
& 18:12:34.00 $-$09:30:02.6 & IrS & 58688.27 & 2219 & 2.9 & $0.174 \pm 0.015$ & 4.1 & 0.23 \\
& 18:12:33.98 $-$09:30:02.5 & IrS & 58937.57 & 2468 & 5.2 & $0.137 \pm 0.018$  & 3.8 & 0.32 \\
 \hline
 $151228$ & 14:16:03.76 $-$17:39:56.0 & IrS & 58654.05 & 1270 & 6.2 & $0.200 \pm 0.013$  & 0.12 
 & 0.10
 \\
  & 14:16:03.76 $-$17:39:55.8 & IrS & 58681.12 & 1297 & 2.9 & $0.445 \pm 0.027$  & 0.36 
  & 0.10
 \\
 & 14:16:03.73 $-$17:39:55.4 & IrS & 58990.08 & 1606 & 6.2 & $0.196 \pm 0.014$  & 0.84 
 & 0.43
 \\
 \hline
 $170112$ & 01:00:54.06 $-$17:12:43.6 & IrS & 58572.72 & 808 & 6.2 & $0.147 \pm 0.020$  & 13
 & 0.17
 \\
 & 01:00:54.06 $-$17:12:43.5 & IrS & 58619.58 & 855 & 2.8 & $0.241 \pm 0.030$  & 13
 & 0.43
 \\
 & 01:00:54.07 $-$17:12:43.7 & IrS & 58906.98 & 1142 & 6.3 & $0.151 \pm 0.026$  & 13 & 0.46 \\
\hline
\end{tabular}
\end{footnotesize}\end{center}\end{table*}

\section{VLA follow-up observations, data reduction, and source identification}
\label{VLAobs}
Observations of the GRBs in our sample were carried out using the VLA in its C and B configurations. We observed all GRBs in both C- and S-bands (overall frequency range of 1.0-8.0\,GHz). A summary of our observations is reported in Table \ref{tb:TableRMS}. Data calibration, imaging, source identification and related measurements were carried out following a procedure similar to what is described in \citet{Artkop2019}, which we briefly summarize in what follows. Results of our analysis are reported in Table \ref{tb:Table}.

The raw data were calibrated using the VLA automated calibration pipeline in CASA \citep{McMullin2007}. After the initial calibration, the data were further inspected and flagged as needed for residual RFI. Imaging was carried out using the CASA \textit{TCLEAN} algorithm in interactive mode. All cleaned images were visually inspected to identify sources with signal-to-noise ratio (SNR) $\gtrsim5$. Source maximum flux densities and positions were calculated using \textit{IMSTAT} within a circular region of radius comparable to the Full Width at Half Maximum (FWHM) of the nominal synthesized beam, centered around the location of the source as identified through visual inspection. Peak flux density errors were calculated as the quadrature sum of the image RMS noise at the source location (which accounts for the telescope primary beam correction) and a systematic absolute flux calibration error. The last was assumed to be of 5\% for all observations that used 3C286 as flux calibrator \citep{Perley2017}, and of 10\% for all observations that made use of 3C48 as flux calibrator (given the on-going flaring behavior of this source). 

Position errors were calculated by dividing the clean beam semi-major axis (as derived using the \textit{IMFIT} task in CASA) by the SNR of the source. For sources with low SNR, position errors estimated using this procedure are more conservative than position errors returned by \textit{IMFIT} (via elliptical Gaussian fits), while for sources with SNR$\gtrsim 10$ the two methods are found to closely agree  \citep[to better than the estimated systematic position uncertainty, conservatively estimated to be of $\approx 0.1\arcsec$;][]{Helfand2015, Palliyaguru2016,Artkop2019}. When the position error calculated as above is found to be smaller than the VLA systematic position uncertainty, we set the position error equal to this systematic uncertainty. 

For all VLA sources identified in our images within the GRBs error regions, we searched for previously known radio sources co-located with them. To this end, we queried the catalogs by the National VLA Sky Survey \citep[NVSS;][]{Condon1998}, VLA FIRST \citep{Becker1994}, and the NASA/IPAC Extragalactic Database \citep[NED;][]{Helou1995}, searching for the closest previously known radio source found within $1\arcmin$ of each of our VLA sources. We also obtained quick-look images from the VLA Sky Survey \citep[VLASS;][]{Lacey2020} by searching for VLASS fields with phase centers within 1\,deg of the VLA candidate position (see Table \ref{tb:Table}).

We note that the observed distribution of projected physical offsets of short GRBs from their hosts has a median of 5\,kpc \citep{Fong2010}. For the short GRB population as a whole, $\gtrsim 25\%$ have projected offsets of $\lesssim 10$\,kpc; and $\gtrsim 5\%$ have projected offsets of $\gtrsim 20$\,kpc \citep{Fong2010}. Thus, searching for known sources located within $1\,\arcmin$ of the position of our VLA sources (which at distances of  40-200\,Mpc corresponds to a physical offsets of $\sim 10-50$\,kpc) leaves room to find not only coincident counterparts but also potential host galaxies.
\subsection{GRB\,130626}
GRB\,130626 triggered the \textit{Swift}/BAT at 10:51:03 on 26 June 2013  UTC \citep{GCN14931}. The refined analysis of the BAT data enabled the localization of the GRB to R.A.=18:12:30.6 Dec=$-$09:31:29.9 (J2000), with a position uncertainty of $1.8\arcmin$ \citep[90\% containment radius;][]{GCN14942}.

We observed the field of GRB\,130626 with the VLA in its BnA configuration starting at 03:19:34 on 8 July 2019 UT in C-band, and at 05:35:12 on 24 July 2019 UT in S-band. Our observations, including calibration, lasted a total of 1\,hr in each band. The cleaned image central rms was $\approx 7.3 \mu$Jy at 6.2\,GHz and $\approx 12 \mu$Jy at 2.9\,GHz. 
One radio source was identified in both our C-band and S-band images within the $1.8\arcmin$ position uncertainty of the BAT error circle for GRB\,130626 \citep{GCN14942}. No known radio source was found within $1\arcmin$ of such source in NVSS, FIRST, or VLASS, while the closest source reported in NED is an Infrared Source (see Table \ref{tb:Table}).

To test for any variability of the radio counterpart candidate identified in our VLA images of the GRB\,130626 field, we re-observed the field in C-band starting at 13:44:31 on 29 March 2020 UT with the VLA in its C configuration. The clean image central rms was $\approx 15 \mu$Jy at 5.2\,GHz. This second VLA observation was affected by significant RFI which led us to flag all spectral windows with nominal central frequencies above 6.1\,GHz. Based on our two observations of this field in C-band, we conclude that no significant flux density variation is found for the candidate radio counterpart of GRB\,130626 over the timescales of our observations. A comparison of Epoch 1 data with imaging restricted to the same spectral channels that remained unflagged in Epoch 2 confirmed this result (see Table \ref{tb:Table}). 

Using our first C-band observation and the S-band observation of the candidate radio source in the field of GRB\,130626, we derive a spectral index of $\beta=-0.90\pm0.30$, where we adopt the notation $F_{\nu}\propto \nu^{\beta}$. Within the large errors, this spectral index is broadly compatible with optically thin synchrotron emission expected from radio flares of NS mergers \citep{Nakar2011}, as well as with radio emission associated with star-formation in galaxies \citep[$-1.1 \lesssim \beta \lesssim -0.4$;][]{Seymour2008}. This spectral index is also not inconsistent, within the large errors, with $\beta > -0.6$ found for $90\%$ of flat-spectrum AGN \citep[e.g.,][and references therein]{Itoh2020}.

Finally, to test for the potential presence of extended radio emission and gain more information about the morphology of the radio candidate, we also calculate the compactness $C$ defined as $C=F_{\rm 3\,GHz,int}/F_{\rm 3\,GHz, peak}$, i.e. the ratio of the integrated and peak fluxes as measured using the \textit{IMFIT} task in CASA. We find $C=1.038\pm0.089$, fully compatible with the range $0.9<C<1.5$ typically used in the classification of point-like radio sources at this same frequency \citep{Mooley2013}. This also agrees with the fact that an elliptical Gaussian fit of the source using \textit{IMFIT} returns no evidence for extended morphology.

\subsection{GRB\,141205}
GRB\,141205 triggered the \textit{Swift}/BAT at 14:51:45 on 5 December 2015 UT \citep{GCN17137}. The refined analysis of the BAT data enabled the localization of the GRB to R.A.=06:11:26.1, Dec=+37:52:32.2 (J2000), with a position uncertainty of $2.0\arcmin$ \citep[90\% containment radius;][]{GCN17140}.

We observed the field of GRB\,141205 with the VLA in its B configuration starting at 21:19:13 on  2  April  2019 UT in C-band, and at 9:37:42 on 14 May, 2019 UT in S-band. Our observations, including calibration, lasted a total of 1\,hr in each band. The clean image central rms was $\approx 6.5\,\mu$Jy at 6.2\,GHz and $\approx 10\,\mu$Jy at 2.8\,GHz. 
None of sources identified in our VLA images are within the BAT $2.0\arcmin$ error circle for this GRB. 

\subsection{GRB\,151228A}
GRB\,151228 triggered the \textit{Swift}/BAT  at 03:05:12 on 28 December 2015 UT
\citep{GCN18731}. The refined analysis of the BAT data enabled the localization of the GRB to R.A.=14:16:04.0 Dec=-17:39:52.7 (J2000), with a position uncertainty of $1.8\arcmin$ \citep[90\% containment radius;][]{GCN18754}

We observed the field of GRB\,151228 with the VLA in its B configuration starting at 01:10:08 on 20 June 2019 UT in C-band, and in its BnA configuration at 02:46:55 on  17  July 2019 UT in S-band. Our observations, including calibration, lasted a total of 1\,hr in each band. The cleaned image central rms was $\approx 8.6 \mu$Jy at 6.2\,GHz and $\approx 15 \mu$Jy at 2.9\,GHz.  
One source identified in both our C-band and S-band images is found within the $1.8\arcmin$ position uncertainty of the BAT error circle for GRB\,151228 \citep{GCN18754}. No known radio source was found within $1\arcmin$ of such source in NVSS, FIRST, or VLASS, while the closest source reported in NED is an Infrared Source (see Table \ref{tb:Table}).

To test for any variability of this candidate radio counterpart, we re-observed the field of GRB\,151228 in C-band starting at 02:14:25 on 21 May 2020 UT with the VLA in its C configuration. The clean image central rms was $\approx 10 \mu$Jy at 6.2\,GHz. Based on our two observations of this field in C-band, we conclude that no significant flux density variation is found for the identified VLA source over the timescales of our observations (see Table \ref{tb:Table}). 

Using our first C-band observation and the S-band observation of the candidate radio source in the field of GRB\,151228, we derive a spectral index of $\beta=-1.05\pm0.12$. This spectral index is compatible with optically thin synchrotron emission expected from radio flares of NS mergers \citep{Nakar2011}, as well as with radio emission associated with star-formation in galaxies \citep[$-1.1 \lesssim \beta \lesssim -0.4$;][]{Seymour2008}. However, a flat spectrum \citep[$\beta>-0.6$;][]{Itoh2020} AGN is disfavored.

Finally, we find $C=1.21\pm0.19$ for the compactness parameter, fully compatible with the range $0.9<C<1.5$ typically used in the classification of point-like sources at 3\,GHz \citep{Mooley2013}. This also agrees with the fact that an elliptical Gaussian fit of the source using \textit{IMFIT} returns no evidence for extended morphology.

\subsection{GRB\,170112}
GRB\,170112 triggered the \textit{Swift}/BAT at 02:02:00 UTC on 12 January, 2017 UT 
\citep{GCN20436}. The refined analysis of the BAT data enabled the localization of the GRB to R.A.=01:00:55.7, Dec=-17:13:57.9, (J2000), with a position uncertainty of $2.5\arcmin$ \citep[90\% containment radius;][]{GCN20443}.

We observed the field of GRB\,170112 with the VLA in its B configuration starting at 17:28:30 UTC on 30 March 2019, in C-band, and at 13:57:23 on 16 May 2019 in S-band. Our observations, including calibration, lasted a total of 1\,hr in each band. The cleaned image central rms was $\approx 7.9 \mu$Jy at 6.2\,GHz and $\approx 20 \mu$Jy at 2.8\,GHz.  
One source identified in both our C-band and S-band images is found  within  the $2.5\arcmin$ position  uncertainty  of  the BAT error circle for GRB\,170112 \citep{GCN20443}. No known radio source was found within $1\arcmin$ of such source in NVSS, FIRST, or VLASS, while the closest source reported in NED is an Infrared Source (see Table \ref{tb:Table}).

To  test  for  any  variability  of the candidate VLA counterpart in the GRB\,170112 field,  we  re-observed the field in C-band starting at 23:37:08 on 27 February 2020 UT with the VLA in its C configuration.  The clean image central rms was $\approx 20\,\mu$Jy at 6.3\,GHz (with the image being dynamic range limited due to the presence of two bright sources in the field). Based on our two observations of this field in C-band, we conclude that no significant flux density variation is found for the identified VLA source  over  the  timescales  of  our  observations  (see Table \ref{tb:Table}).  

Using our first C-band observation and the S-band observation of the candidate radio source in the field of GRB\,151228, we derive a spectral index of $\beta=-0.62\pm0.18$. This spectral index is compatible with optically thin synchrotron emission expected from radio flares of NS mergers \citep{Nakar2011}, with radio emission associated with star-formation in galaxies \citep[$-1.1 \lesssim \beta \lesssim -0.4$;][]{Seymour2008}, and with emission from a flat-spectrum \citep[$\beta > -0.6$;][]{Itoh2020} AGN.

Finally, we find $C=0.89\pm0.39$ for the compactness parameter, fully compatible with the range $0.9<C<1.5$ typically used in the classification of point-like sources at 3\,GHz \citep{Mooley2013}. This also agrees with the fact that an elliptical Gaussian fit of the source using \textit{IMFIT} returns no evidence for extended morphology.

\begin{figure}
\includegraphics[width=0.51\textwidth]{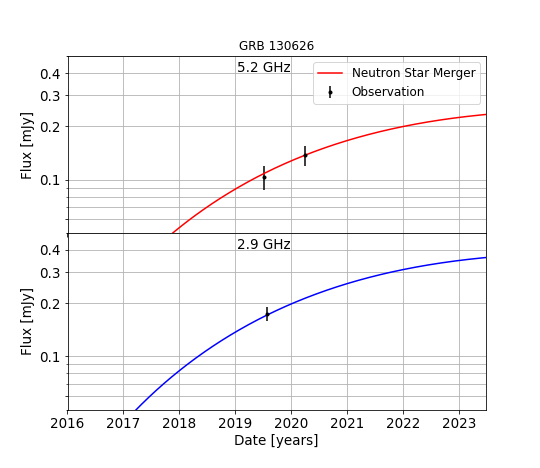}
\centering
\caption{VLA follow-up observations in the direction of GRB\,130626 at 5.2\,GHz (top) and 2.9\,GHz (bottom). We also show a comparison with a NS merger radio flare fit with NS merger time equal to the time of GRB\,130626, ejecta mass similar to that observed for GRB170817, circum-merger medium density of $7\times10^{-2}$\,cm$^{-3}$, and distance of 102\,Mpc (solid lines).\\}
\label{fig:radio}
\end{figure}
%%%%%%%%%%%

\section{Discussion}
\label{results}
Among our sample of four GRBs, three (GRB\,130626, GRB\,151228, and GRB\,170112; see Table \ref{tb:Table}) have a candidate VLA counterpart within their \textit{Swift}/BAT localization region. None of these candidates  showed significant time variability of the flux density at  5-6\,GHz within the timescale of our observations (and our flux density measurement errors), and all of them have compactness $C$ compatible with point sources at 3\,GHz \citep{Mooley2013}. 
No radio source was found within $1\arcmin$ of our radio candidates in NVSS/FIRST/VLASS. We note that VLASS is conducted at 3\,GHz, thus matching the S-band observations of GRBs in our sample. Howeover, the rms values of the VLASS images containing the GRB fields is $\approx 0.17-0.18$\,mJy at 3\,GHz, and thus it is not surprising that no significant radio excess was found around the position of our VLA candidate counterparts (whose 3\,GHz flux is below the VLASS $3\times$rms sensitivity for these fields; see Table \ref{tb:Table}).

As discussed in Section \ref{VLAobs}, the spectral indices of the VLA candidate radio sources in Table \ref{tb:Table} are compatible with both flares of binary NS mergers and star formation in galaxies. For two of our three GRBs, GRB\,130626 and GRB\,170112, the radio candidate has spectral index also compatible with a flat spectrum AGN, within the uncertainties. 

In light of the above, in this Section we discuss the possible origin of the radio candidates identified in our GRB sample. We first quantify the chances of finding persistent and variable radio sources not related to GRB emission within error areas comparable to those of GRBs in our sample. Then, we discuss possible scenarios for the origin of these candidates in terms of AGN or star-formation emission, as well  binary NS merger radio flares.

\subsection{Contaminating persistent and variable radio sources}
Given the lack of significant temporal variability of the measured flux densities at 6\,GHz (within the timescales and flux density measurement errors of our observations), we can estimate the number of persistent and unrelated radio sources one would expect in the \textit{Swift}/BAT error area of GRBs in our sample. 
Using our spectral index measurement, we derive an upper-bound on the 1.4\,GHz emission from these candidates counterparts of $F_{\rm 1.4\,GHz}\lesssim 0.3-1$\,mJy, assuming an optically thin spectrum. 
\citet{Mooley2013} have estimated the average number of persistent radio sources with 1.4\,GHz flux densities $\gtrsim 0.1\,$mJy to be $\approx 910 \,{\rm deg}^{-2}$. The three GRBs for which we found a candidate VLA counterpart have 90\% containment error radii of $ 1.8-2.5\,\arcmin$, implying expected average numbers of \textit{persistent} radio sources in their error regions of $\approx 2.6-5$ per GRB. Thus, our results are consistent with potentially un-associated persistent radio sources in the GRB fields. 

We note, however, that because we have collected only two epochs for each GRB in our sample, our assessment of lack of variability is restricted to poorly sampled timescales. Further follow-up could clarify whether the sources we have identified are truly persistent. As noted in \citet{Mooley2013}, the variable/transient radio sky is a lot quieter than the persistent radio sky, with an average areal sky density of variable radio sources about 1\% that of persistent radio sources. Thus, if the candidate radio sources identified here were found to be variable in future follow-up observations, the likelihood of a spurious association with the corresponding GRBs would be  reduced, as $\approx 0.26-0.5$ variable sources would be expected within the error circles of these GRBs. Consequently, the Poisson probability of finding one or more unrelated variable radio sources within the GRB error area would be $20-40\%$ for GRBs in our sample. 

\subsection{Testing an AGN or star-formation origin}
Star-forming galaxies and AGNs are likely to be the two major populations of radio (cm) sources that could contribute false positives to a search for binary NS merger radio flares when dealing with relatively large localization areas \citep[e.g.,][]{Condon1992,Sadler1999,Smolcic2008,Baran2016,Palliyaguru2016}. Below $\approx 200\,\mu$Jy at 3\,GHz, star-forming galaxies start to dominate in terms of fractional contribution to the total source sample \citep{Baran2016}, although low-luminosity AGNs may also be present \citep{Mooley2013}. At the mJy level, the transient/variable radio sky is dominated by AGNs \citep{Sadler1999}. 

The compactness parameters and \textit{IMFIT} size measurements for the VLA candidate radio sources found in our analysis do not provide clear evidence for extended emission, and suggest a point-like morphology within the VLA S-band resolution of $\approx 2.7-3.3\,\arcsec$ (major axis FWHM of the synthesized beam in B configuration). We can compare this constraint with effective radii of short GRB host galaxies in the cosmological sample presented by \citet{Fong2013}. Those have effective radii in the range $0.2\arcsec-1.2\arcsec$, or $\approx 3.5$\,kpc in terms of median physical size. This median size would correspond to an angular radius of $\gtrsim 3.9\arcsec$ at a distances $\lesssim 200$\,Mpc. Thus, if the radio candidates found in the GRBs error regions were the host galaxies of the GRB themselves, our measurements would suggest distances $\gtrsim 200$\,Mpc to be more likely.

If all of the radio emission of the VLA candidates is associated with synchrotron radiation from relativistic electrons accelerated by supernovae in a normal galaxy, the implied star formation rate (SFR) can be estimated as \citep{Murphy2011,Perley2013}:
\begin{equation}
   \left( \frac{SFR}{\rm M_{\odot}\,{\rm yr}^{-1}}\right)=6.35\times10^{-29}\left(\frac{L_{\rm 1.4\,GHz}}{\rm erg\,s^{-1} Hz^{-1}}\right).
\end{equation}
If we assume that the VLA candidates are associated with SFR in galaxies located at 200\,Mpc, the implied luminosity densities would be in the range  $L_{1.4\,\rm GHz}\lesssim (1.6-4.6)\times10^{28}$\,erg\,s$^{-1}$\,Hz$^{-1}$, and SFRs in the range $SFR\lesssim (1-2.8)\,{\rm M_{\odot}\,{\rm yr}^{-1}}$, compatible with normal galaxies. This is also compatible with SFR rates of cosmological short GRB host galaxies, which are found to be in the range $(0.2-6)\,{\rm M_{\odot}\,{\rm yr}^{-1}}$. The host galaxy of GW170817, NGC4993, had an uncommonly low SFR of $0.01\,{\rm M_{\odot}\,{\rm yr}^{-1}}$ \citep{Fong2017}.

On the other hand, if we assume the radio emission originated from a typical short GRB host galaxy at the median short GRB redshift of $z=0.72$ \citep{Berger2014}, the 1.4\,GHz flux densities estimated above would imply luminosity densities of $L_{1.4\,\rm GHz}\lesssim (7.9-20.3)\times10^{30}$\,erg\,s$^{-1}$\,Hz$^{-1}$ (or, equivalently, radio power values of log$(L_{\rm 1.4\,GHz}[{\rm W/Hz}])\lesssim 23.9-24.3$).  Thus, under the cosmological GRB host galaxy hypothesis, and under the assumption of optically thin spectral indices down to 1.4\,GHz, our measured flux densities would favor an AGN origin of the radio emission.

As a way to further differentiate between AGNs and normal galaxies, we can use \textit{WISE} color information \citep{Mateos2012} about our VLA candidates. 
Among our three VLA candidates, only the one associated with GRB\,130626 has $W1$, $W2$, $W3$ colors and corresponding errors available in the \textit{WISE} catalog. For this candidate we derive $W1-W2=-0.047\pm0.035$ and $W2-W3=1.04\pm0.12$, which places this candidate out of the AGN wedge defined by $W2–W3>2.517$ and $W1-W2> 0.315 \times (W2-W3) - 0.222$ \citep{Mateos2012}.

\subsection{Testing a binary NS merger origin}
We finally consider whether the VLA candidates identified in the error region of the GRBs in our sample could be consistent, in terms of flux densities in the 3-6\,GHz range, with a NS merger radio flare origin. 

To this end, we adopt an ejecta model similar to that observed for the case of GW170817. Specifically, we assume a two-component ejecta, with component masses $0.04$\,M$_\odot$ and $0.01$\,M$_\odot$, and speed of $0.1$\,c and $0.3$\,c, respectively \citep{Bartos2019}. We then simulate the expected radio light curves for a range of circum-merger densities \citep[$n_{\rm ISM}=10^{-4}-10$\,cm$^{-3}$;][]{Berger2014} 
and distances (40-200\,Mpc), and identify the density and distance parameters corresponding to the radio flare light curve that best matches the observed data. 

We find that two of our three radio sources (those identified in the error regions of GRB\,151228 and GRB\,170112) cannot be adequately fit with this model. On the other hand, the observations corresponding to GRB\,130626 can be fit with plausible model parameters. We show the data and the best fit light curves for the two observed frequencies in Figure \ref{fig:radio}. This best fit model corresponds to a circum-merger density of $7\times10^{-2}$\,cm$^{-3}$ and a distance of $102$\,Mpc. 

\section{Summary and Conclusion}
\label{sec:conclusion}
We have carried out VLA observations of a sample of four \textit{Swift} short GRBs for which no afterglow was identified. Our observations aimed at discovering potential slowly-evolving radio counterparts from nearby binary NS mergers. For three out of the four GRBs in our sample we have found a corresponding VLA source not contained in previous radio catalogs.  

Our VLA candidates did not show significant flux density variability within the limited time range (and flux density errors) of our observations. This is consistent with slowly-varying radio emission from a NS merger, or unrelated persistent radio sources. The fact that we found radio candidates within the GRB error areas is also consistent with expectations from the sky density of persistent radio sources.

The morphology of our radio candidates as estimated via the compactness parameter at 3\,GHz disfavors an origin related to star formation in galaxies similar to that of short GRB hosts within 200\,Mpc. On the other hand, the measured radio flux densities and spectral indices imply that an association with galaxies at a redshift equal to the median short GRB redshift would favor an AGN origin for the observed radio flux. For GRB\,130626, available WISE observations do not support an AGN origin. However, an origin related to star formation in a galaxy at a distance larger than 200\,Mpc (but smaller than the median short GRB redshift), with size similar to that of cosmological short GRB host galaxies, is plausible.

Finally, by  fitting  a NS  merger  radio  flare  model  on  our  observations,  we found that the radio source in the error region of GRB\,130626 is consistent with the radio flare of a NS merger similar to GW170817, with circum-merger density of $7\times10^{-2}$\,cm$^{-3}$ at a distance of 102\,Mpc.
Further radio observations will be able to provide additional discrimination between the above scenarios for GRB\,130626.  

In conclusion we note that under the optimistic assumptions that up to 10\% of short GRBs in the known sample could be nearby \citep{Gupte2018}, the Poisson probability of detecting at least one of them in a sample of 4 GRBs is $\approx 33$\%, so searches like the ones here presented should continue in a more systematic fashion. Targeting samples of $\gtrsim 23$ GRBs would bring us to $\gtrsim 90$\% chances of detecting at least one event. However, given the possible contamination from unrelated persistent radio sources,  the follow-up campaign should be extended to several years so as to probe significant flux density variability. We note that these types of radio follow-ups can be conducted on the VLA in filler-time mode, thus they are relatively inexpensive.

\acknowledgements
The National Radio Astronomy Observatory is a facility of the National Science Foundation operated under cooperative agreement by Associated Universities, Inc. A.B., A.C., C.G., H.H., and P.R. acknowledge support from the National Science Foundation via the CAREER award \#1455090 and award PHY-1907975. J.M. acknowledges support from the Clark Scholars Program and Lubbock High School. I.B. acknowledges the support of the National Science Foundation under grant PHY-1911796 and the Alfred P. Sloan Foundation. Zs.M. and Sz.M. are grateful to National Science Foundation under grant PHY-1708028 and Columbia University in the City of New York for their generous support.

\bibliography{bibliography.bib}

\begin{thebibliography}{}
\expandafter\ifx\csname natexlab\endcsname\relax\def\natexlab#1{#1}\fi

\bibitem[{{Abbott} {et~al.}(2017{\natexlab{a}}){Abbott}, {Abbott}, {Abbott},
  {Acernese}, {Ackley}, {Adams}, {Adams}, {Addesso}, {Adhikari}, {Adya},
  {Affeldt}, {Afrough}, {Agarwal}, {Agathos}, {Agatsuma}, {Aggarwal}, {Aguiar},
  {Aiello}, {Ain}, {Ajith}, {Allen}, {Allen}, {Allocca}, {Aloy}, {Altin},
  {Amato}, {Ananyeva}, {Anderson}, {Anderson}, {Angelova}, {Antier}, {Appert},
  {Arai}, {Araya}, {Areeda}, {Arnaud}, {Arun}, {Ascenzi}, {Ashton}, {Ast},
  {Aston}, {Astone}, {Atallah}, {Aufmuth}, {Aulbert}, {AultONeal}, {Austin}, \&
  {Avila-Alvarez}}]{Abbott2017b}
{Abbott}, B.~P., {Abbott}, R., {Abbott}, T.~D., {et~al.} 2017{\natexlab{a}},
  \apjl, 848, L13

\bibitem[{{Abbott} {et~al.}(2017{\natexlab{b}}){Abbott}, {Abbott}, {Abbott},
  {Acernese}, {Ackley}, {Adams}, {Adams}, {Addesso}, {Adhikari}, {Adya},
  {Affeldt}, {Afrough}, {Agarwal}, {Agathos}, {Agatsuma}, {Aggarwal}, {Aguiar},
  {Aiello}, {Ain}, {Ajith}, {Allen}, {Allen}, {Allocca}, {Altin}, {Amato},
  {Ananyeva}, {Anderson}, {Anderson}, {Angelova}, {Antier}, {Appert}, {Arai},
  {Araya}, {Areeda}, {Arnaud}, {Arun}, {Ascenzi}, {Ashton}, {Ast}, {Aston},
  {Astone}, {Atallah}, {Aufmuth}, {Aulbert}, {AultONeal}, {Austin},
  {Avila-Alvarez}, {et~al.}}]{Abbott2017a}
---. 2017{\natexlab{b}}, \prl, 119, 161101

\bibitem[{{Abbott} {et~al.}(2017{\natexlab{c}}){Abbott}, {Abbott}, {Abbott},
  {Acernese}, {Ackley}, {Adams}, {Adams}, {Addesso}, {Adhikari}, {Adya},
  {Affeldt}, {Afrough}, {Agarwal}, {Agathos}, {Agatsuma}, {Aggarwal}, {Aguiar},
  {Aiello}, {Ain}, {Ajith}, {Allen}, {Allen}, {Allocca}, {Altin}, {Amato},
  {Ananyeva}, {Anderson}, {Anderson}, {Angelova}, {Antier}, {Appert}, {Arai},
  {Araya}, {Areeda}, {Arnaud}, {Arun}, {Ascenzi}, {Ashton}, {Ast}, {Aston},
  {Astone}, {Atallah}, {Aufmuth}, {Aulbert}, {AultONeal}, {Austin},
  {Avila-Alvarez}, {et~al.}}]{Abbott2017c}
---. 2017{\natexlab{c}}, \apjl, 848, L12

\bibitem[{{Abbott}(2018)}]{Abbott2018}
{Abbott}, B.~P. e.~a. 2018, Living Reviews in Relativity, 21, 3

\bibitem[{{Alexander} {et~al.}(2017){Alexander}, {Berger}, {Fong}, {Williams},
  {Guidorzi}, {Margutti}, {Metzger}, {Annis}, {Blanchard}, {Brout}, {Brown},
  {Chen}, {Chornock}, {Cowperthwaite}, {Drout}, {Eftekhari}, {Frieman}, {Holz},
  {Nicholl}, {Rest}, {Sako}, {Soares-Santos}, \& {Villar}}]{Alexander2017}
{Alexander}, K.~D., {Berger}, E., {Fong}, W., {et~al.} 2017, \apjl, 848, L21

\bibitem[{{Arcavi} {et~al.}(2017){Arcavi}, {Hosseinzadeh}, {Howell}, {McCully},
  {Poznanski}, {Kasen}, {Barnes}, {Zaltzman}, {Vasylyev}, {Maoz}, \&
  {Valenti}}]{Arcavi2017}
{Arcavi}, I., {Hosseinzadeh}, G., {Howell}, D.~A., {et~al.} 2017, \nat, 551, 64

\bibitem[{{Artkop} {et~al.}(2019){Artkop}, {Smith}, {Corsi}, {Giacintucci},
  {Peters}, {Perna}, {Cenko}, \& {Clarke}}]{Artkop2019}
{Artkop}, K., {Smith}, R., {Corsi}, A., {et~al.} 2019, \apj, 884, 16

\bibitem[{{Baran} {et~al.}(2016){Baran}, {Smolcic}, {Delvecchio}, {Novak},
  {Delhaize}, {Laigle}, {Ilbert}, \& {(Vla-)Cosmos Collaboration}}]{Baran2016}
{Baran}, N., {Smolcic}, V., {Delvecchio}, I., {et~al.} 2016, in Active Galactic
  Nuclei: What's in a Name?, 15

\bibitem[{{Barthelmy} {et~al.}(2015){Barthelmy}, {Cummings}, {Gehrels},
  {et~al.}}]{GCN18754}
{Barthelmy}, S., {Cummings}, J., {Gehrels}, N., {et~al.} 2015, GRB Coordinate
  Network, 18754

\bibitem[{{Bartos} {et~al.}(2019){Bartos}, {Lee}, {Corsi}, {M{\'a}rka}, \&
  {M{\'a}rka}}]{Bartos2019}
{Bartos}, I., {Lee}, K.~H., {Corsi}, A., {M{\'a}rka}, Z., \& {M{\'a}rka}, S.
  2019, \mnras, 485, 4150

\bibitem[{{Becker} {et~al.}(1994){Becker}, {White}, \& {Helfand}}]{Becker1994}
{Becker}, R.~H., {White}, R.~L., \& {Helfand}, D.~J. 1994, in Astronomical
  Society of the Pacific Conference Series, Vol.~61, Astronomical Data Analysis
  Software and Systems III, ed. D.~R. {Crabtree}, R.~J. {Hanisch}, \&
  J.~{Barnes}, 165

\bibitem[{{Berger}(2014)}]{Berger2014}
{Berger}, E. 2014, \araa, 52, 43

\bibitem[{{Blanchard} {et~al.}(2017){Blanchard}, {Berger}, {Fong}, {Nicholl},
  {Leja}, {Conroy}, {Alexander}, {Margutti}, {Williams}, {Doctor}, {Chornock},
  {Villar}, {Cowperthwaite}, {Annis}, {Brout}, {Brown}, {Chen}, {Eftekhari},
  {Frieman}, {Holz}, {Metzger}, {Rest}, {Sako}, \&
  {Soares-Santos}}]{Blanchard2017}
{Blanchard}, P.~K., {Berger}, E., {Fong}, W., {et~al.} 2017, \apjl, 848, L22

\bibitem[{{Chornock} {et~al.}(2017){Chornock}, {Berger}, {Kasen},
  {Cowperthwaite}, {Nicholl}, {Villar}, {Alexand er}, {Blanchard}, {Eftekhari},
  {Fong}, {Margutti}, {Williams}, {Annis}, {Brout}, {Brown}, {Chen}, {Drout},
  {Farr}, {Foley}, {Frieman}, {Fryer}, {Herner}, {Holz}, {Kessler}, {Matheson},
  {Metzger}, {Quataert}, {Rest}, {Sako}, {Scolnic}, {Smith}, \&
  {Soares-Santos}}]{Chornock2017}
{Chornock}, R., {Berger}, E., {Kasen}, D., {et~al.} 2017, \apjl, 848, L19

\bibitem[{{Condon}(1992)}]{Condon1992}
{Condon}, J.~J. 1992, \araa, 30, 575

\bibitem[{{Condon} {et~al.}(1998){Condon}, {Cotton}, {Greisen}, {Yin},
  {Perley}, {Taylor}, \& {Broderick}}]{Condon1998}
{Condon}, J.~J., {Cotton}, W.~D., {Greisen}, E.~W., {et~al.} 1998, \aj, 115,
  1693

\bibitem[{{Corsi} {et~al.}(2018){Corsi}, {Hallinan}, {Lazzati}, {Mooley},
  {Murphy}, {Frail}, {Carbone}, {Kaplan}, {Murphy}, {Kulkarni}, \&
  {Hotokezaka}}]{Corsi2018}
{Corsi}, A., {Hallinan}, G.~W., {Lazzati}, D., {et~al.} 2018, \apjl, 861, L10

\bibitem[{{Coulter} {et~al.}(2017){Coulter}, {Foley}, {Kilpatrick}, {Drout},
  {Piro}, {Shappee}, {Siebert}, {Simon}, {Ulloa}, {Kasen}, {Madore},
  {Murguia-Berthier}, {Pan}, {Prochaska}, {Ramirez-Ruiz}, {Rest}, \&
  {Rojas-Bravo}}]{Coulter2017}
{Coulter}, D.~A., {Foley}, R.~J., {Kilpatrick}, C.~D., {et~al.} 2017, Science,
  358, 1556

\bibitem[{{Cowperthwaite} {et~al.}(2017){Cowperthwaite}, {Berger}, {Villar},
  {Metzger}, {Nicholl}, {Chornock}, {Blanchard}, {Fong}, {Margutti},
  {Soares-Santos}, {Alexander}, {Allam}, {Annis}, {Brout}, {Brown}, {Butler},
  {Chen}, {Diehl}, {Doctor}, {Drout}, {Eftekhari}, {Farr}, {Finley}, {Foley},
  {Frieman}, {Fryer}, {Garc{\'\i}a-Bellido}, {Gill}, {Guillochon}, {Herner},
  {Holz}, {Kasen}, {Kessler}, {Marriner}, {Matheson}, {Neilsen}, {Quataert},
  {Palmese}, {Rest}, {Sako}, {Scolnic}, {Smith}, {Tucker}, {Williams},
  {Balbinot}, {Carlin}, {Cook}, {Durret}, {Li}, {Lopes}, {Louren{\c{c}}o},
  {Marshall}, {Medina}, {Muir}, {Mu{\~n}oz}, {Sauseda}, {Schlegel}, {Secco},
  {Vivas}, {Wester}, {Zenteno}, {Zhang}, {Abbott}, {Banerji}, {Bechtol},
  {Benoit-L{\'e}vy}, {Bertin}, {Buckley-Geer}, {Burke}, {Capozzi}, {Carnero
  Rosell}, {Carrasco Kind}, {Castander}, {Crocce}, {Cunha}, {D'Andrea}, {da
  Costa}, {Davis}, {DePoy}, {Desai}, {Dietrich}, {Drlica-Wagner}, {Eifler},
  {Evrard}, {Fernand ez}, {Flaugher}, {Fosalba}, {Gaztanaga}, {Gerdes},
  {Giannantonio}, {Goldstein}, {Gruen}, {Gruendl}, {Gutierrez}, {Honscheid},
  {Jain}, {James}, {Jeltema}, {Johnson}, {Johnson}, {Kent}, {Krause}, {Kron},
  {Kuehn}, {Nuropatkin}, {Lahav}, {Lima}, {Lin}, {Maia}, {March}, {Martini},
  {McMahon}, {Menanteau}, {Miller}, {Miquel}, {Mohr}, {Neilsen}, {Nichol},
  {Ogando}, {Plazas}, {Roe}, {Romer}, {Roodman}, {Rykoff}, {Sanchez},
  {Scarpine}, {Schindler}, {Schubnell}, {Sevilla-Noarbe}, {Smith}, {Smith},
  {Sobreira}, {Suchyta}, {Swanson}, {Tarle}, {Thomas}, {Thomas}, {Troxel},
  {Vikram}, {Walker}, {Wechsler}, {Weller}, {Yanny}, \&
  {Zuntz}}]{Cowperthwaite2017}
{Cowperthwaite}, P.~S., {Berger}, E., {Villar}, V.~A., {et~al.} 2017, \apjl,
  848, L17

\bibitem[{{Cummings} {et~al.}(2014)}]{GCN17137}
{Cummings}, J., {et~al.} 2014, GRB Coordinate Network, 17137

\bibitem[{{De Pasquale} {et~al.}(2013){De Pasquale}, {Barthelmy}, {Burrows},
  {et~al.}}]{GCN14931}
{De Pasquale}, M., {Barthelmy}, S., {Burrows}, D., {et~al.} 2013, GRB
  Coordinate Network, 14931

\bibitem[{{Dichiara} {et~al.}(2020){Dichiara}, {Troja}, {O'Connor}, {Marshall},
  {Beniamini}, {Cannizzo}, {Lien}, \& {Sakamoto}}]{Dichiara2020}
{Dichiara}, S., {Troja}, E., {O'Connor}, B., {et~al.} 2020, \mnras, 492, 5011

\bibitem[{{Dobie} {et~al.}(2018){Dobie}, {Kaplan}, {Murphy}, {Lenc}, {Mooley},
  {Lynch}, {Corsi}, {Frail}, {Kasliwal}, \& {Hallinan}}]{Dobie2018}
{Dobie}, D., {Kaplan}, D.~L., {Murphy}, T., {et~al.} 2018, \apjl, 858, L15

\bibitem[{{Drout} {et~al.}(2017){Drout}, {Piro}, {Shappee}, {Kilpatrick},
  {Simon}, {Contreras}, {Coulter}, {Foley}, {Siebert}, {Morrell}, {Boutsia},
  {Di Mille}, {Holoien}, {Kasen}, {Kollmeier}, {Madore}, {Monson},
  {Murguia-Berthier}, {Pan}, {Prochaska}, {Ramirez-Ruiz}, {Rest}, {Adams},
  {Alatalo}, {Ba{\~n}ados}, {Baughman}, {Beers}, {Bernstein}, {Bitsakis},
  {Campillay}, {Hansen}, {Higgs}, {Ji}, {Maravelias}, {Marshall}, {Moni Bidin},
  {Prieto}, {Rasmussen}, {Rojas-Bravo}, {Strom}, {Ulloa},
  {Vargas-Gonz{\'a}lez}, {Wan}, \& {Whitten}}]{Drout2017}
{Drout}, M.~R., {Piro}, A.~L., {Shappee}, B.~J., {et~al.} 2017, Science, 358,
  1570

\bibitem[{{Fong} \& {Berger}(2013)}]{Fong2013}
{Fong}, W., \& {Berger}, E. 2013, \apj, 776, 18

\bibitem[{{Fong} {et~al.}(2010){Fong}, {Berger}, \& {Fox}}]{Fong2010}
{Fong}, W., {Berger}, E., \& {Fox}, D.~B. 2010, \apj, 708, 9

\bibitem[{{Fong} {et~al.}(2016){Fong}, {Metzger}, {Berger}, \&
  {{\"O}zel}}]{Fong2016}
{Fong}, W., {Metzger}, B.~D., {Berger}, E., \& {{\"O}zel}, F. 2016, \apj, 831,
  141

\bibitem[{{Fong} {et~al.}(2017){Fong}, {Berger}, {Blanchard}, {Margutti},
  {Cowperthwaite}, {Chornock}, {Alexander}, {Metzger}, {Villar}, {Nicholl},
  {Eftekhari}, {Williams}, {Annis}, {Brout}, {Brown}, {Chen}, {Doctor},
  {Diehl}, {Holz}, {Rest}, {Sako}, \& {Soares-Santos}}]{Fong2017}
{Fong}, W., {Berger}, E., {Blanchard}, P.~K., {et~al.} 2017, \apjl, 848, L23

\bibitem[{{Gehrels} {et~al.}(2004){Gehrels}, {Chincarini}, {Giommi}, {Mason},
  {Nousek}, {Wells}, {White}, {Barthelmy}, {Burrows}, {Cominsky}, {Hurley},
  {Marshall}, {M{\'e}sz{\'a}ros}, {Roming}, {Angelini}, {Barbier}, {Belloni},
  {Campana}, {Caraveo}, {Chester}, {Citterio}, {Cline}, {Cropper}, {Cummings},
  {Dean}, {Feigelson}, {Fenimore}, {Frail}, {Fruchter}, {Garmire}, {Gendreau},
  {Ghisellini}, {Greiner}, {Hill}, {Hunsberger}, {Krimm}, {Kulkarni}, {Kumar},
  {Lebrun}, {Lloyd-Ronning}, {Markwardt}, {Mattson}, {Mushotzky}, {Norris},
  {Osborne}, {Paczynski}, {Palmer}, {Park}, {Parsons}, {Paul}, {Rees},
  {Reynolds}, {Rhoads}, {Sasseen}, {Schaefer}, {Short}, {Smale}, {Smith},
  {Stella}, {Tagliaferri}, {Takahashi}, {Tashiro}, {Townsley}, {Tueller},
  {Turner}, {Vietri}, {Voges}, {Ward}, {Willingale}, {Zerbi}, \&
  {Zhang}}]{Gehrels2004}
{Gehrels}, N., {Chincarini}, G., {Giommi}, P., {et~al.} 2004, \apj, 611, 1005

\bibitem[{{Ghirlanda} {et~al.}(2019){Ghirlanda}, {Salafia}, {Paragi},
  {Giroletti}, {Yang}, {Marcote}, {Blanchard}, {Agudo}, {An}, {Bernardini},
  {Beswick}, {Branchesi}, {Campana}, {Casadio}, {Chassand e-Mottin}, {Colpi},
  {Covino}, {D'Avanzo}, {D'Elia}, {Frey}, {Gawronski}, {Ghisellini}, {Gurvits},
  {Jonker}, {van Langevelde}, {Melandri}, {Moldon}, {Nava}, {Perego},
  {Perez-Torres}, {Reynolds}, {Salvaterra}, {Tagliaferri}, {Venturi},
  {Vergani}, \& {Zhang}}]{Ghirlanda2019}
{Ghirlanda}, G., {Salafia}, O.~S., {Paragi}, Z., {et~al.} 2019, Science, 363,
  968

\bibitem[{Goldstein {et~al.}(2017)Goldstein, Veres, Burns, Briggs, Hamburg,
  Kocevski, Wilson-Hodge, Preece, Poolakkil, Roberts, Hui, Connaughton,
  Racusin, von Kienlin, Canton, Christensen, Littenberg, Siellez, Blackburn,
  Broida, Bissaldi, Cleveland, Gibby, Giles, Kippen, McBreen, McEnery, Meegan,
  Paciesas, \& Stanbro}]{Goldstein2017}
Goldstein, A., Veres, P., Burns, E., {et~al.} 2017, The Astrophysical Journal,
  848, L14

\bibitem[{{Granot} {et~al.}(2018){Granot}, {Gill}, {Guetta}, \& {De
  Colle}}]{Granot2018}
{Granot}, J., {Gill}, R., {Guetta}, D., \& {De Colle}, F. 2018, \mnras, 481,
  1597

\bibitem[{{Gupte} \& {Bartos}(2018)}]{Gupte2018}
{Gupte}, N., \& {Bartos}, I. 2018, arXiv e-prints, arXiv:1808.06238

\bibitem[{{Haggard} {et~al.}(2017){Haggard}, {Nynka}, {Ruan}, {Kalogera},
  {Cenko}, {Evans}, \& {Kennea}}]{Haggard2017}
{Haggard}, D., {Nynka}, M., {Ruan}, J.~J., {et~al.} 2017, \apjl, 848, L25

\bibitem[{{Hajela} {et~al.}(2019){Hajela}, {Margutti}, {Alexander},
  {Kathirgamaraju}, {Baldeschi}, {Guidorzi}, {Giannios}, {Fong}, {Wu},
  {MacFadyen}, {Paggi}, {Berger}, {Blanchard}, {Chornock}, {Coppejans},
  {Cowperthwaite}, {Eftekhari}, {Gomez}, {Hosseinzadeh}, {Laskar}, {Metzger},
  {Nicholl}, {Paterson}, {Radice}, {Sironi}, {Terreran}, {Villar}, {Williams},
  {Xie}, \& {Zrake}}]{Hajela2019}
{Hajela}, A., {Margutti}, R., {Alexander}, K.~D., {et~al.} 2019, \apjl, 886,
  L17

\bibitem[{{Hallinan} {et~al.}(2017){Hallinan}, {Corsi}, {Mooley}, {Hotokezaka},
  {Nakar}, {Kasliwal}, {Kaplan}, {Frail}, {Myers}, {Murphy}, {De}, {Dobie},
  {Allison}, {Bannister}, {Bhalerao}, {Chandra}, {Clarke}, {Giacintucci}, {Ho},
  {Horesh}, {Kassim}, {Kulkarni}, {Lenc}, {Lockman}, {Lynch}, {Nichols},
  {Nissanke}, {Palliyaguru}, {Peters}, {Piran}, {Rana}, {Sadler}, \&
  {Singer}}]{Hallinan2017}
{Hallinan}, G., {Corsi}, A., {Mooley}, K.~P., {et~al.} 2017, Science, 358, 1579

\bibitem[{{Helfand} {et~al.}(2015){Helfand}, {White}, \&
  {Becker}}]{Helfand2015}
{Helfand}, D.~J., {White}, R.~L., \& {Becker}, R.~H. 2015, \apj, 801, 26

\bibitem[{{Helou} {et~al.}(1995){Helou}, {Madore}, {Schmitz}, {Wu}, {Corwin},
  {Lague}, {Bennett}, \& {Sun}}]{Helou1995}
{Helou}, G., {Madore}, B.~F., {Schmitz}, M., {et~al.} 1995, {The NASA/IPAC
  Extragalactic Database}, ed. D.~{Egret} \& M.~A. {Albrecht}, Vol. 203, 95

\bibitem[{{Hotokezaka} {et~al.}(2013){Hotokezaka}, {Kiuchi}, {Kyutoku},
  {Okawa}, {Sekiguchi}, {Shibata}, \& {Taniguchi}}]{Hotokezaka2013}
{Hotokezaka}, K., {Kiuchi}, K., {Kyutoku}, K., {et~al.} 2013, \prd, 87, 024001

\bibitem[{{Hotokezaka} {et~al.}(2018){Hotokezaka}, {Kiuchi}, {Shibata},
  {Nakar}, \& {Piran}}]{Hotokezaka2018}
{Hotokezaka}, K., {Kiuchi}, K., {Shibata}, M., {Nakar}, E., \& {Piran}, T.
  2018, \apj, 867, 95

\bibitem[{{Hotokezaka} {et~al.}(2016){Hotokezaka}, {Nissanke}, {Hallinan},
  {Lazio}, {Nakar}, \& {Piran}}]{Hotokezaka2016}
{Hotokezaka}, K., {Nissanke}, S., {Hallinan}, G., {et~al.} 2016, \apj, 831, 190

\bibitem[{{Itoh} {et~al.}(2020){Itoh}, {Utsumi}, {Inoue}, {Ohta}, {Doi},
  {Morokuma}, {Kawabata}, \& {Tanaka}}]{Itoh2020}
{Itoh}, R., {Utsumi}, Y., {Inoue}, Y., {et~al.} 2020, arXiv e-prints,
  arXiv:2008.00038

\bibitem[{{Kasliwal} {et~al.}(2017){Kasliwal}, {Nakar}, {Singer}, {Kaplan},
  {Cook}, {Van Sistine}, {Lau}, {Fremling}, {Gottlieb}, {Jencson}, {Adams},
  {Feindt}, {Hotokezaka}, {Ghosh}, {Perley}, {Yu}, {Piran}, {Allison},
  {Anupama}, {Balasubramanian}, {Bannister}, {Bally}, {Barnes}, {Barway},
  {Bellm}, {Bhalerao}, {Bhattacharya}, {Blagorodnova}, {Bloom}, {Brady},
  {Cannella}, {Chatterjee}, {Cenko}, {Cobb}, {Copperwheat}, {Corsi}, {De},
  {Dobie}, {Emery}, {Evans}, {Fox}, {Frail}, {Frohmaier}, {Goobar}, {Hallinan},
  {Harrison}, {Helou}, {Hinderer}, {Ho}, {Horesh}, {Ip}, {Itoh}, {Kasen},
  {Kim}, {Kuin}, {Kupfer}, {Lynch}, {Madsen}, {Mazzali}, {Miller}, {Mooley},
  {Murphy}, {Ngeow}, {Nichols}, {Nissanke}, {Nugent}, {Ofek}, {Qi}, {Quimby},
  {Rosswog}, {Rusu}, {Sadler}, {Schmidt}, {Sollerman}, {Steele}, {Williamson},
  {Xu}, {Yan}, {Yatsu}, {Zhang}, \& {Zhao}}]{Kasliwal2017}
{Kasliwal}, M.~M., {Nakar}, E., {Singer}, L.~P., {et~al.} 2017, Science, 358,
  1559

\bibitem[{{Klose} {et~al.}(2019){Klose}, {Nicuesa Guelbenzu}, {Micha{\l}owski},
  {Hunt}, {Hartmann}, {Greiner}, {Rossi}, {Palazzi}, \& {Bernuzzi}}]{Klose2019}
{Klose}, S., {Nicuesa Guelbenzu}, A.~M., {Micha{\l}owski}, M.~J., {et~al.}
  2019, \apj, 887, 206

\bibitem[{{Kouveliotou} {et~al.}(1993){Kouveliotou}, {Meegan}, {Fishman},
  {Bhat}, {Briggs}, {Koshut}, {Paciesas}, \& {Pendleton}}]{Kouveliotou1993}
{Kouveliotou}, C., {Meegan}, C.~A., {Fishman}, G.~J., {et~al.} 1993, \apjl,
  413, L101

\bibitem[{{Lacy} {et~al.}(2020){Lacy}, {Baum}, {Chandler}, {Chatterjee},
  {Clarke}, {Deustua}, {English}, {Farnes}, {Gaensler}, {Gugliucci},
  {Hallinan}, {Kent}, {Kimball}, {Law}, {Lazio}, {Marvil}, {Mao}, {Medlin},
  {Mooley}, {Murphy}, {Myers}, {Osten}, {Richards}, {Rosolowsky}, {Rudnick},
  {Schinzel}, {Sivakoff}, {Sjouwerman}, {Taylor}, {White}, {Wrobel},
  {Andernach}, {Beasley}, {Berger}, {Bhatnager}, {Birkinshaw}, {Bower},
  {Brandt}, {Brown}, {Burke-Spolaor}, {Butler}, {Comerford}, {Demorest}, {Fu},
  {Giacintucci}, {Golap}, {G{\"u}th}, {Hales}, {Hiriart}, {Hodge}, {Horesh},
  {Ivezi{\'c}}, {Jarvis}, {Kamble}, {Kassim}, {Liu}, {Loinard}, {Lyons},
  {Masters}, {Mezcua}, {Moellenbrock}, {Mroczkowski}, {Nyland},
  {O{\textquoteright}Dea}, {O{\textquoteright}Sullivan}, {Peters}, {Radford},
  {Rao}, {Robnett}, {Salcido}, {Shen}, {Sobotka}, {Witz}, {Vaccari}, {van
  Weeren}, {Vargas}, {Williams}, \& {Yoon}}]{Lacey2020}
{Lacy}, M., {Baum}, S.~A., {Chandler}, C.~J., {et~al.} 2020, \pasp, 132, 035001

\bibitem[{{Law} {et~al.}(2018){Law}, {Gaensler}, {Metzger}, {Ofek}, \&
  {Sironi}}]{Law2018}
{Law}, C.~J., {Gaensler}, B.~M., {Metzger}, B.~D., {Ofek}, E.~O., \& {Sironi},
  L. 2018, \apjl, 866, L22

\bibitem[{{Lazzati} {et~al.}(2018){Lazzati}, {Perna}, {Morsony},
  {Lopez-Camara}, {Cantiello}, {Ciolfi}, {Giacomazzo}, \&
  {Workman}}]{Lazzati2018}
{Lazzati}, D., {Perna}, R., {Morsony}, B.~J., {et~al.} 2018, \prl, 120, 241103

\bibitem[{{Lee} {et~al.}(2020){Lee}, {Bartos}, {Privon}, {Rose}, \&
  {Torrey}}]{Lee2020}
{Lee}, K.~H., {Bartos}, I., {Privon}, G.~C., {Rose}, J.~C., \& {Torrey}, P.
  2020, arXiv e-prints, arXiv:2007.00563

\bibitem[{{Lien} {et~al.}(2017){Lien}, {Barthelmy}, {Cummings},
  {et~al.}}]{GCN20443}
{Lien}, A., {Barthelmy}, S., {Cummings}, J., {et~al.} 2017, GRB Coordinate
  Network, 20436

\bibitem[{{Lien} {et~al.}(2016){Lien}, {Sakamoto}, {Barthelmy}, {Baumgartner},
  {Cannizzo}, {Chen}, {Collins}, {Cummings}, {Gehrels}, {Krimm}, {Markwardt},
  {Palmer}, {Stamatikos}, {Troja}, \& {Ukwatta}}]{Lien2016}
{Lien}, A., {Sakamoto}, T., {Barthelmy}, S.~D., {et~al.} 2016, \apj, 829, 7

\bibitem[{{Liu} {et~al.}(2020){Liu}, {Gao}, \& {Zhang}}]{Liu2020}
{Liu}, L.-D., {Gao}, H., \& {Zhang}, B. 2020, \apj, 890, 102

\bibitem[{{Makhathini} {et~al.}(2020){Makhathini}, {Mooley}, {Brightman},
  {Hotokezaka}, {Nayana}, {Intema}, {Dobie}, {Lenc}, {Perley}, {Fremling},
  {Moldon}, {Lazzati}, {Kaplan}, {Balasubramanian}, {Brown}, {Carbone},
  {Chandra}, {Corsi}, {Camilo}, {Deller}, {Frail}, {Murphy}, {Murphy}, {Nakar},
  {Smirnov}, {Beswick}, {Fender}, {Hallinan}, {Heywood}, {Kasliwal}, {Lee},
  {Lu}, {Rana}, {Perkins}, {White}, {Jozsa}, {Hugo}, \&
  {Kamphuis}}]{Makhathini2020}
{Makhathini}, S., {Mooley}, K.~P., {Brightman}, M., {et~al.} 2020, arXiv
  e-prints, arXiv:2006.02382

\bibitem[{{Mandhai} {et~al.}(2018){Mandhai}, {Tanvir}, {Lamb}, {Levan}, \&
  {Tsang}}]{Mandhai2018}
{Mandhai}, S., {Tanvir}, N., {Lamb}, G., {Levan}, A., \& {Tsang}, D. 2018,
  Galaxies, 6, 130

\bibitem[{{Margalit} \& {Piran}(2020)}]{Margalit2020}
{Margalit}, B., \& {Piran}, T. 2020, \mnras, 495, 4981

\bibitem[{{Margutti} {et~al.}(2017){Margutti}, {Berger}, {Fong}, {Guidorzi},
  {Alexander}, {Metzger}, {Blanchard}, {Cowperthwaite}, {Chornock},
  {Eftekhari}, {Nicholl}, {Villar}, {Williams}, {Annis}, {Brown}, {Chen},
  {Doctor}, {Frieman}, {Holz}, {Sako}, \& {Soares-Santos}}]{Margutti2017}
{Margutti}, R., {Berger}, E., {Fong}, W., {et~al.} 2017, \apjl, 848, L20

\bibitem[{{Margutti} {et~al.}(2018){Margutti}, {Alexander}, {Xie}, {Sironi},
  {Metzger}, {Kathirgamaraju}, {Fong}, {Blanchard}, {Berger}, {MacFadyen},
  {Giannios}, {Guidorzi}, {Hajela}, {Chornock}, {Cowperthwaite}, {Eftekhari},
  {Nicholl}, {Villar}, {Williams}, \& {Zrake}}]{Margutti2018}
{Margutti}, R., {Alexander}, K.~D., {Xie}, X., {et~al.} 2018, \apjl, 856, L18

\bibitem[{{Markwardt} {et~al.}(2014){Markwardt}, {Barthelmy}, {Baumgartner},
  {et~al.}}]{GCN17140}
{Markwardt}, C., {Barthelmy}, S., {Baumgartner}, W., {et~al.} 2014, GRB
  Coordinate Network, 17140

\bibitem[{{Mateos} {et~al.}(2012){Mateos}, {Alonso-Herrero}, {Carrera},
  {Blain}, {Watson}, {Barcons}, {Braito}, {Severgnini}, {Donley}, \&
  {Stern}}]{Mateos2012}
{Mateos}, S., {Alonso-Herrero}, A., {Carrera}, F.~J., {et~al.} 2012, \mnras,
  426, 3271

\bibitem[{{Matsumoto} \& {Piran}(2020)}]{Matsumoto2020}
{Matsumoto}, T., \& {Piran}, T. 2020, \mnras, 492, 4283

\bibitem[{{McMullin} {et~al.}(2007){McMullin}, {Waters}, {Schiebel}, {Young},
  \& {Golap}}]{McMullin2007}
{McMullin}, J.~P., {Waters}, B., {Schiebel}, D., {Young}, W., \& {Golap}, K.
  2007, in Astronomical Society of the Pacific Conference Series, Vol. 376,
  Astronomical Data Analysis Software and Systems XVI, ed. R.~A. {Shaw},
  F.~{Hill}, \& D.~J. {Bell}, 127

\bibitem[{{Metzger} \& {Fern{\'a}ndez}(2014)}]{Metzger2014}
{Metzger}, B.~D., \& {Fern{\'a}ndez}, R. 2014, \mnras, 441, 3444

\bibitem[{{Mingo} {et~al.}(2017){Mingo}, {Burrows}, {Gronwall},
  {et~al.}}]{GCN20436}
{Mingo}, B., {Burrows}, D., {Gronwall}, C., {et~al.} 2017, GRB Coordinate
  Network, 20436

\bibitem[{{Mooley} {et~al.}(2013){Mooley}, {Frail}, {Ofek}, {Miller},
  {Kulkarni}, \& {Horesh}}]{Mooley2013}
{Mooley}, K.~P., {Frail}, D.~A., {Ofek}, E.~O., {et~al.} 2013, \apj, 768, 165

\bibitem[{{Mooley} {et~al.}(2018{\natexlab{a}}){Mooley}, {Nakar}, {Hotokezaka},
  {Hallinan}, {Corsi}, {Frail}, {Horesh}, {Murphy}, {Lenc}, {Kaplan}, {de},
  {Dobie}, {Chand ra}, {Deller}, {Gottlieb}, {Kasliwal}, {Kulkarni}, {Myers},
  {Nissanke}, {Piran}, {Lynch}, {Bhalerao}, {Bourke}, {Bannister}, \&
  {Singer}}]{Mooley2018a}
{Mooley}, K.~P., {Nakar}, E., {Hotokezaka}, K., {et~al.} 2018{\natexlab{a}},
  \nat, 554, 207

\bibitem[{{Mooley} {et~al.}(2018{\natexlab{b}}){Mooley}, {Frail}, {Dobie},
  {Lenc}, {Corsi}, {De}, {Nayana}, {Makhathini}, {Heywood}, {Murphy}, {Kaplan},
  {Chandra}, {Smirnov}, {Nakar}, {Hallinan}, {Camilo}, {Fender}, {Goedhart},
  {Groot}, {Kasliwal}, {Kulkarni}, \& {Woudt}}]{Mooley2018b}
{Mooley}, K.~P., {Frail}, D.~A., {Dobie}, D., {et~al.} 2018{\natexlab{b}},
  \apjl, 868, L11

\bibitem[{{Mooley} {et~al.}(2018{\natexlab{c}}){Mooley}, {Deller}, {Gottlieb},
  {Nakar}, {Hallinan}, {Bourke}, {Frail}, {Horesh}, {Corsi}, \&
  {Hotokezaka}}]{Mooley2018c}
{Mooley}, K.~P., {Deller}, A.~T., {Gottlieb}, O., {et~al.} 2018{\natexlab{c}},
  \nat, 561, 355

\bibitem[{{Murphy} {et~al.}(2011){Murphy}, {Condon}, {Schinnerer}, {Kennicutt},
  {Calzetti}, {Armus}, {Helou}, {Turner}, {Aniano}, {Beir{\~a}o}, {Bolatto},
  {Brandl}, {Croxall}, {Dale}, {Donovan Meyer}, {Draine}, {Engelbracht},
  {Hunt}, {Hao}, {Koda}, {Roussel}, {Skibba}, \& {Smith}}]{Murphy2011}
{Murphy}, E.~J., {Condon}, J.~J., {Schinnerer}, E., {et~al.} 2011, \apj, 737,
  67

\bibitem[{{Nakar}(2019)}]{Nakar2019}
{Nakar}, E. 2019, arXiv e-prints, arXiv:1912.05659

\bibitem[{{Nakar} \& {Piran}(2011)}]{Nakar2011}
{Nakar}, E., \& {Piran}, T. 2011, \nat, 478, 82

\bibitem[{{Nicholl} {et~al.}(2017){Nicholl}, {Berger}, {Kasen}, {Metzger},
  {Elias}, {Brice{\~n}o}, {Alexander}, {Blanchard}, {Chornock},
  {Cowperthwaite}, {Eftekhari}, {Fong}, {Margutti}, {Villar}, {Williams},
  {Brown}, {Annis}, {Bahramian}, {Brout}, {Brown}, {Chen}, {Clemens},
  {Dennihy}, {Dunlap}, {Holz}, {Marchesini}, {Massaro}, {Moskowitz},
  {Pelisoli}, {Rest}, {Ricci}, {Sako}, {Soares-Santos}, \&
  {Strader}}]{Nicholl2017}
{Nicholl}, M., {Berger}, E., {Kasen}, D., {et~al.} 2017, \apjl, 848, L18

\bibitem[{{Palliyaguru} {et~al.}(2016){Palliyaguru}, {Corsi}, {Kasliwal},
  {Cenko}, {Frail}, {Perley}, {Mishra}, {Singer}, {Gal-Yam}, {Nugent}, \&
  {Surace}}]{Palliyaguru2016}
{Palliyaguru}, N.~T., {Corsi}, A., {Kasliwal}, M.~M., {et~al.} 2016, \apjl,
  829, L28

\bibitem[{{Perley} \& {Perley}(2013)}]{Perley2013}
{Perley}, D.~A., \& {Perley}, R.~A. 2013, \apj, 778, 172

\bibitem[{Perley \& Butler(2017)}]{Perley2017}
Perley, R.~A., \& Butler, B.~J. 2017, The Astrophysical Journal Supplement
  Series, 230, 7

\bibitem[{{Ricci} {et~al.}(2020){Ricci}, {Troja}, {Bruni}, {Matsumoto}, {Piro},
  {O'Connor}, {Piran}, {Navaieelavasani}, {Corsi}, {Giacomazzo}, \&
  {Wieringa}}]{Ricci2020}
{Ricci}, R., {Troja}, E., {Bruni}, G., {et~al.} 2020, arXiv e-prints,
  arXiv:2008.03659

\bibitem[{{Sadler} {et~al.}(1999){Sadler}, {McIntyre}, {Jackson}, \&
  {Cannon}}]{Sadler1999}
{Sadler}, E.~M., {McIntyre}, V.~J., {Jackson}, C.~A., \& {Cannon}, R.~D. 1999,
  \pasa, 16, 247

\bibitem[{{Sakamoto} {et~al.}(2013){Sakamoto}, {Barthelmy}, {Baumgartner},
  {et~al.}}]{GCN14942}
{Sakamoto}, T., {Barthelmy}, S., {Baumgartner}, W., {et~al.} 2013, GRB
  Coordinate Network, 14942

\bibitem[{{Savchenko} {et~al.}(2017){Savchenko}, {Ferrigno}, {Kuulkers},
  {Bazzano}, {Bozzo}, {Brandt}, {Chenevez}, {Courvoisier}, {Diehl}, {Domingo},
  {Hanlon}, {Jourdain}, {von Kienlin}, {Laurent}, {Lebrun}, {Lutovinov},
  {Martin-Carrillo}, {Mereghetti}, {Natalucci}, {Rodi}, {Roques}, {Sunyaev}, \&
  {Ubertini}}]{Savchenko2017}
{Savchenko}, V., {Ferrigno}, C., {Kuulkers}, E., {et~al.} 2017, \apjl, 848, L15

\bibitem[{{Schroeder} {et~al.}(2020){Schroeder}, {Margalit}, {Fong}, {Metzger},
  {Williams}, {Paterson}, {Alexander}, {Laskar}, {Goyal}, \&
  {Berger}}]{Schroeder2020}
{Schroeder}, G., {Margalit}, B., {Fong}, W.-f., {et~al.} 2020, arXiv e-prints,
  arXiv:2006.07434

\bibitem[{{Seymour} {et~al.}(2008){Seymour}, {Dwelly}, {Moss}, {McHardy},
  {Zoghbi}, {Rieke}, {Page}, {Hopkins}, \& {Loaring}}]{Seymour2008}
{Seymour}, N., {Dwelly}, T., {Moss}, D., {et~al.} 2008, \mnras, 386, 1695

\bibitem[{{Smol{\v{c}}i{\'c}} {et~al.}(2008){Smol{\v{c}}i{\'c}}, {Schinnerer},
  {Scodeggio}, {Franzetti}, {Aussel}, {Bondi}, {Brusa}, {Carilli}, {Capak},
  {Charlot}, {Ciliegi}, {Ilbert}, {Ivezi{\'c}}, {Jahnke}, {McCracken},
  {Obri{\'c}}, {Salvato}, {Sand ers}, {Scoville}, {Trump}, {Tremonti}, {Tasca},
  {Walcher}, \& {Zamorani}}]{Smolcic2008}
{Smol{\v{c}}i{\'c}}, V., {Schinnerer}, E., {Scodeggio}, M., {et~al.} 2008,
  \apjs, 177, 14

\bibitem[{{Soares-Santos} {et~al.}(2017){Soares-Santos}, {Holz}, {Annis},
  {Chornock}, {Herner}, {Berger}, {Brout}, {Chen}, {Kessler}, {Sako}, {Allam},
  {Tucker}, {Butler}, {Palmese}, {Doctor}, {Diehl}, {Frieman}, {Yanny}, {Lin},
  {Scolnic}, {Cowperthwaite}, {Neilsen}, {Marriner}, {Kuropatkin}, {Hartley},
  {Paz-Chinch{\'o}n}, {Alexander}, {Balbinot}, {Blanchard}, {Brown}, {Carlin},
  {Conselice}, {Cook}, {Drlica-Wagner}, {Drout}, {Durret}, {Eftekhari}, {Farr},
  {Finley}, {Foley}, {Fong}, {Fryer}, {Garc{\'\i}a-Bellido}, {Gill}, {Gruendl},
  {Hanna}, {Kasen}, {Li}, {Lopes}, {Louren{\c{c}}o}, {Margutti}, {Marshall},
  {Matheson}, {Medina}, {Metzger}, {Mu{\~n}oz}, {Muir}, {Nicholl}, {Quataert},
  {Rest}, {Sauseda}, {Schlegel}, {Secco}, {Sobreira}, {Stebbins}, {Villar},
  {Vivas}, {Walker}, {Wester}, {Williams}, {Zenteno}, {Zhang}, {Abbott},
  {Abdalla}, {Banerji}, {Bechtol}, {Benoit-L{\'e}vy}, {Bertin}, {Brooks},
  {Buckley-Geer}, {Burke}, {Carnero Rosell}, {Carrasco Kind}, {Carretero},
  {Castander}, {Crocce}, {Cunha}, {D'Andrea}, {da Costa}, {Davis}, {Desai},
  {Dietrich}, {Doel}, {Eifler}, {Fernand ez}, {Flaugher}, {Fosalba},
  {Gaztanaga}, {Gerdes}, {Giannantonio}, {Goldstein}, {Gruen}, {Gschwend},
  {Gutierrez}, {Honscheid}, {Jain}, {James}, {Jeltema}, {Johnson}, {Johnson},
  {Kent}, {Krause}, {Kron}, {Kuehn}, {Kuhlmann}, {Lahav}, {Lima}, {Maia},
  {March}, {McMahon}, {Menanteau}, {Miquel}, {Mohr}, {Nichol}, {Nord}, {Ogand
  o}, {Petravick}, {Plazas}, {Romer}, {Roodman}, {Rykoff}, {Sanchez},
  {Scarpine}, {Schubnell}, {Sevilla-Noarbe}, {Smith}, {Smith}, {Suchyta},
  {Swanson}, {Tarle}, {Thomas}, {Thomas}, {Troxel}, {Vikram}, {Wechsler},
  {Weller}, {Dark Energy Survey}, \& {Dark Energy Camera GW-EM
  Collaboration}}]{Soares-Santos2017}
{Soares-Santos}, M., {Holz}, D.~E., {Annis}, J., {et~al.} 2017, \apjl, 848, L16

\bibitem[{{Troja} {et~al.}(2017){Troja}, {Piro}, {van Eerten}, {Wollaeger},
  {Im}, {Fox}, {Butler}, {Cenko}, {Sakamoto}, {Fryer}, {Ricci}, {Lien}, {Ryan},
  {Korobkin}, {Lee}, {Burgess}, {Lee}, {Watson}, {Choi}, {Covino}, {D'Avanzo},
  {Fontes}, {Gonz{\'a}lez}, {Khandrika}, {Kim}, {Kim}, {Lee}, {Lee}, {Kutyrev},
  {Lim}, {S{\'a}nchez-Ram{\'\i}rez}, {Veilleux}, {Wieringa}, \&
  {Yoon}}]{Troja2017}
{Troja}, E., {Piro}, L., {van Eerten}, H., {et~al.} 2017, \nat, 551, 71

\bibitem[{{Troja} {et~al.}(2018{\natexlab{a}}){Troja}, {Ryan}, {Piro}, {van
  Eerten}, {Cenko}, {Yoon}, {Lee}, {Im}, {Sakamoto}, {Gatkine}, {Kutyrev}, \&
  {Veilleux}}]{Troja2018b}
{Troja}, E., {Ryan}, G., {Piro}, L., {et~al.} 2018{\natexlab{a}}, Nature
  Communications, 9, 4089

\bibitem[{{Troja} {et~al.}(2018{\natexlab{b}}){Troja}, {Piro}, {Ryan}, {van
  Eerten}, {Ricci}, {Wieringa}, {Lotti}, {Sakamoto}, \& {Cenko}}]{Troja2018a}
{Troja}, E., {Piro}, L., {Ryan}, G., {et~al.} 2018{\natexlab{b}}, \mnras, 478,
  L18

\bibitem[{{Troja} {et~al.}(2019){Troja}, {van Eerten}, {Ryan}, {Ricci},
  {Burgess}, {Wieringa}, {Piro}, {Cenko}, \& {Sakamoto}}]{Troja2019}
{Troja}, E., {van Eerten}, H., {Ryan}, G., {et~al.} 2019, \mnras, 489, 1919

\bibitem[{{Ukwatta} {et~al.}(2015){Ukwatta}, {Barthelmy}, {Cummings},
  {et~al.}}]{GCN18731}
{Ukwatta}, T., {Barthelmy}, S., {Cummings}, J., {et~al.} 2015, GRB Coordinate
  Network, 18731

\bibitem[{{Valenti} {et~al.}(2017){Valenti}, {Sand}, {Yang}, {Cappellaro},
  {Tartaglia}, {Corsi}, {Jha}, {Reichart}, {Haislip}, \&
  {Kouprianov}}]{Valenti2017}
{Valenti}, S., {Sand}, D.~J., {Yang}, S., {et~al.} 2017, \apjl, 848, L24

\bibitem[{{van Eerten} \& {MacFadyen}(2012)}]{vanEerten2012}
{van Eerten}, H.~J., \& {MacFadyen}, A.~I. 2012, \apj, 751, 155

\bibitem[{{von Kienlin} {et~al.}(2019){von Kienlin}, {Veres}, {Roberts},
  {Hamburg}, {Bissaldi}, {Briggs}, {Burns}, {Goldstein}, {Kocevski}, {Preece},
  {Wilson-Hodge}, {Hui}, {Mailyan}, \& {Malacaria}}]{vonKienlin2019}
{von Kienlin}, A., {Veres}, P., {Roberts}, O.~J., {et~al.} 2019, \apj, 876, 89

\bibitem[{{Yue} {et~al.}(2018){Yue}, {Hu}, {Zhang}, {Liang}, {Jin}, {Zou},
  {Fan}, \& {Wei}}]{Yue2018}
{Yue}, C., {Hu}, Q., {Zhang}, F.-W., {et~al.} 2018, \apjl, 853, L10

\end{thebibliography}

\end{document}